\documentclass[preprint]{aastex631}

\usepackage{amsmath}
\usepackage{amssymb}
\usepackage{array,multirow}
\usepackage{comment}
\usepackage[inline]{enumitem}
\usepackage{soul}
\usepackage{svg}
\usepackage{tikz}
\usetikzlibrary{arrows,matrix}
\usepackage{url}

\newcommand{\phimu}{\!\left(\mu,\phi\right)}
\newcommand{\lm}{_{\ell}^{m}}

\submitjournal{AAS Journals}

\shorttitle{Resurfacing Venus}
\shortauthors{Adams \& Laughlin}

\graphicspath{{./}{figures/}}

\begin{document}

\title{Turning Earth into Venus: A Stochastic Model of Possible Evolutions of Terrestrial Topography}

\correspondingauthor{Arthur D. Adams}
\email{arthura@ucr.edu}
\author[0000-0002-7139-3695]{Arthur D. Adams}
\affiliation{Department of Earth and Planetary Sciences \\
University of California, Riverside \\
900 University Ave., Riverside, CA 92507, USA}

\author[0000-0002-3253-2621]{Gregory P. Laughlin}
\affiliation{Department of Astronomy \\ Yale University \\ New Haven, CT 06520}

\begin{abstract}

Venus may have had both an Earth-like climate as well as extensive water oceans and active (or incipient) plate tectonics for an extended interval of its history. The topographical power spectrum of Venus provides important clues to the planet's past evolution. By drawing detailed contrast with the strong low-order odd-$l$ dominated global topography of Earth, we demonstrate that the relatively flat Venusian topography can be interpreted to have arisen from the transition from active terrestrial-like plate tectonics to the current stagnant lid configuration at a time $\tau = 544^{+886}_{-193}$ million years before present. This scenario is plausible if loss of oceans and the attendant transition to a CO$_2$-dominated atmosphere were accompanied by rapid continental-scale erosion, followed by gradual lava resurfacing at an outflow rate $\sim$ 1 km$^{3}$ yr$^{-1}$. We study Venus' proposed topographical relaxation with a global diffusion-like model that adopts terrestrial erosion rates scaled to account for the increased rainfall and temperatures that would accompany a planet-wide transition from an Earth-like climate to the runaway greenhouse climate that could ultimately yield present-day Venus, with an estimate of $5.1^{+1.8}_{-1.1}$ Myr if the global erosion operated as efficiently as that of a typical bedrock river basin on Earth.

\end{abstract}


\section{Introduction} \label{sec:intro}
A potentially fruitful line of inquiry proceeds from a detailed contrasting of the topographic power spectra of Earth and Venus --- two similar-sized planets whose surface features and near-surface environments are strongly divergent. Among the worlds of the Solar System, Earth has been most thoroughly studied using approaches that draw on harmonic analysis \citep[e.g.][]{Vening-Meinesz1951,Balmino1973,Turcotte1987,Rapp1989,Balmino1993,Bills1995a,Gagnon2006}. Other bodies, however, including Venus \citep[e.g.][]{Bills1985a,Konopliv1993,Konopliv1999, McNamee1993,Rappaport1999}, and Mars \citep[e.g.][]{Bills1978,Bills1993,Bills1995b} have also received significant attention.  In this respect, a particularly salient contrast can be drawn by observing that on Venus, the correlation between the local gravity and the topography spectrum is much closer than on Earth \citep[see specifically][]{Bills1985a,Bills1985b,Turcotte1987,McNamee1993}, particularly at large spatial scales. This implies that the large-scale topography on Venus is much more supported by isostatic compensation than on Earth, where the boundaries of tectonic plates contribute significantly to the topographic spectrum.

These differences are equally clear when comparing the surface elevation maps. Whereas Venus exhibits a nearly unimodal elevation distribution, with roughly 60\% of the surface within 500 m of the mean \citep[e.g.][]{Head1981}, the Earth's topography is dominated at large scales by the contrast between continents and oceans, which leads to a strong bimodality in the elevation distribution between the thick continental crust and the relatively thin oceanic crust \citep[e.g.][]{Harrison1981,Whitehead2017,Smrekar2018}. Indeed, much of Earth's coarse topographic structure can be directly attributed to the effects of plate tectonics. Following the original proposal for continental drift by Wegener in 1912, foundational works such as \citet{Wilson1965} put forth the idea that oceanic ridges are interconnected, and their spreading \citep[as detailed in][]{Dietz1961,Hess1962} defines the boundaries and motions of a set of plates that span the globe. In this picture, the observed plate motions are produced by ongoing convection in the lithosphere \citep{McKenzie1969,McKenzie1973}. It is now well established that new plates can be created and destroyed at spreading centers, which are linked to dynamics in the crust and upper mantle \citep[e.g.][]{Buck2015}. While the exact nature of the relation between plate and mantle dynamics is still subject to debate \citep{Wessel2007}, they are intrinsically coupled, with plate tectonics often being framed as the ``surface expression'' of mantle convection \citep[e.g.][]{Bercovici2003,Foley2016}.

Earth's plate tectonics might have taken shape as early as $\sim\!4$ Gya in the form of proto-subduction \citep[e.g.][]{vanHunen2012,Korenaga2013}. Fully global tectonics, however, are thought to have begun in earnest by $\sim\!3.0$--2.7 Gya \citep[e.g.][]{Condie2008,Shirey2008,Shirey2011}, and appear to have provided a stabilizing influence on the global climate. Moreover, it is believed that plate tectonics are made possible on Earth through an active carbon cycle that tempers the climate, and that the temperance is itself maintained partly through active tectonics that aid the carbon cycle, leading to a coupled system between the climate and tectonic activity \citep[e.g.][]{Sleep1995,Driscoll2013,Foley2016}. Indeed, it is now widely posited that active plate tectonics significantly aid planetary habitability \citep[e.g.][]{Bercovici2015b,Stern2016,Ramirez2018}, although the mechanism might not be strictly required if effective carbon cycling is feasible without plates \citep[e.g.][]{Foley2018}.

Venus has neither a temperate climate nor active plate tectonics, and elucidating the history of Venus's tectonic activity constitutes an active area of research. Venus's current crust-mantle configuration appears to be in an inactive lid state, often characterized as a ``stagnant lid'', in contrast with Earth's tectonics which are an example of an active (a.k.a.~mobile) lid state. In a stagnant lid state, there is effectively a single global plate, and the majority of heat flux is released through episodes of volcanism and non-plate tectonic deformation \citep[e.g.][]{Phillips1986,Turcotte1995,Solomatov1996,Basilevsky1998,Bjonnes2012,Smrekar2018,Stern2018a}. To date, however, it is not known whether the current observed inactive lid is the result of a long-term, stable stagnant lid configuration, or instead a relatively quiescent period within an ``episodic'' lid state where mantle and surface overturn events punctuate extended epochs of stasis \citep[e.g.][]{Moresi1998,Weller2015,Rolf2018,Weller2020,Byrne2021}. 

Regardless of the history of Venus's tectonics, Venus's topography has been reshaped by a combination of cratering and volcanic resurfacing \citep[e.g.][]{Strom1994,Hauck1998,Reese1999,Romeo2010,Stein2010,King2018}, with the timescale for resurfacing falling in the range $\sim\!750$--250 Mya \citep[e.g.][]{Schaber1992,Turcotte1993,Head1994,Strom1994,Bottke2015}. And, just as Earth's climate and mantle dynamics are coupled, Venus's current atmosphere is thought to be the product of volcanic degassing from the mantle \citep[e.g.][]{Gillmann2014}.

Assuming that Venus once harbored an Earth-like environment with topography dominated by continents and ocean basins, can we date the onset of the current inactive lid period by modeling the effects that the transition would have on the global topography? As we show, this general plan looks feasible, albeit with the critical requirement that erosion associated with a relatively brief epoch of ocean evaporation occurred at a high rate.

The layout for this article runs as follows. In \S \ref{sec:data} we document the topographic data we use in our analysis.  We develop a  spectral model to reshape Earth's surface via diffusion and volcanism, with the goal of reproducing a Venusian hypsometry in \S \ref{sec:model}. The results of the parameter fitting, as well as the corresponding estimates of physical timescales, are described in \S \ref{sec:results}. We discuss these outcomes in context of past and recent analyses of erosive and volcanic processes on both Earth and Venus in \S \ref{sec:discussion}.

\section{Topographic Data and Elevation Models}\label{sec:data} 
To perform our analyses we rely on digital elevation models (DEMs), which generally factor in the precise shape of the geoid, yielding as close to a purely hypsometric dataset as possible. The datasets for Earth, Venus, and Mars are described as follows.

\begin{description}
    \item [Modern Earth] The ETOPO1 global relief model \citep{Amante2009} uses a combination of topography and bathymetry from bedrock and ice surface data. The latitude and longitude coordinates are referenced from WGS84 \citep{WGS84}, which constitutes an oblate spheroidal model. The elevations are referenced to the geoid, which is equivalent to sea level. The minimum resolution is $1^\prime$ ($\sim 1.8\,{\rm km}$). We note that newer composite models have taken into account improved topography over oceans and ice sheets \citep{Hirt2015}; however, for the purposes of our global-focused analysis the standard ETOPO1 model is more than sufficient, especially when comparing our results for modern Earth with current state-of-art resolutions for Earth's historical configurations.
    \item[Past Earth] \citet{Scotese2018} have published an extensive set of elevation maps at 1-degree resolution that document the surface effects of plate tectonics in 5 Myr intervals back through the Phanerozoic (540 Mya), which we use in our analysis in \S \ref{sec:results}. Briefly speaking, the construction of the PaleoDEMs relies on a global plate tectonic framework with a collection of lithofacies to infer past climatic regions (e.g. seas, deserts, mountain ranges, etc.). Where lithofacial techniques are infeasible, one can use reconstructions from local tectonic histories. This allows for the removal of recent volcanic events, the restoration of ancient mountain ranges, and the recreation of the bathymetry of past oceans.
    \item [Venus] Global topography at a limiting resolution of 5 km/pixel comes from Magellan altimetry \citep{VenusDEM}. The elevations are given relative to a sphere of radius 6051 km. These data have been used in recent analyses of Venus's gravity/topography using harmonic modes, such as \citet{King2018}.
    \item [Mars] The Mars Orbiter Laser Altimeter \cite[MOLA, see][]{Smith2001} provides global topography at resolutions just under $0.5^\prime$ (463 m/pixel). The resulting DEM \citep{Neumann2001,Neumann2003,MarsDEM} is given relative to the areoid \citep{Lemoine2001}, Mars's equivalent of the geoid. A more recent model of the surface gravitational potential is described in \citet{Gorski2018}.
\end{description}

\begin{figure}
\begin{center}
\includegraphics[width=17cm]{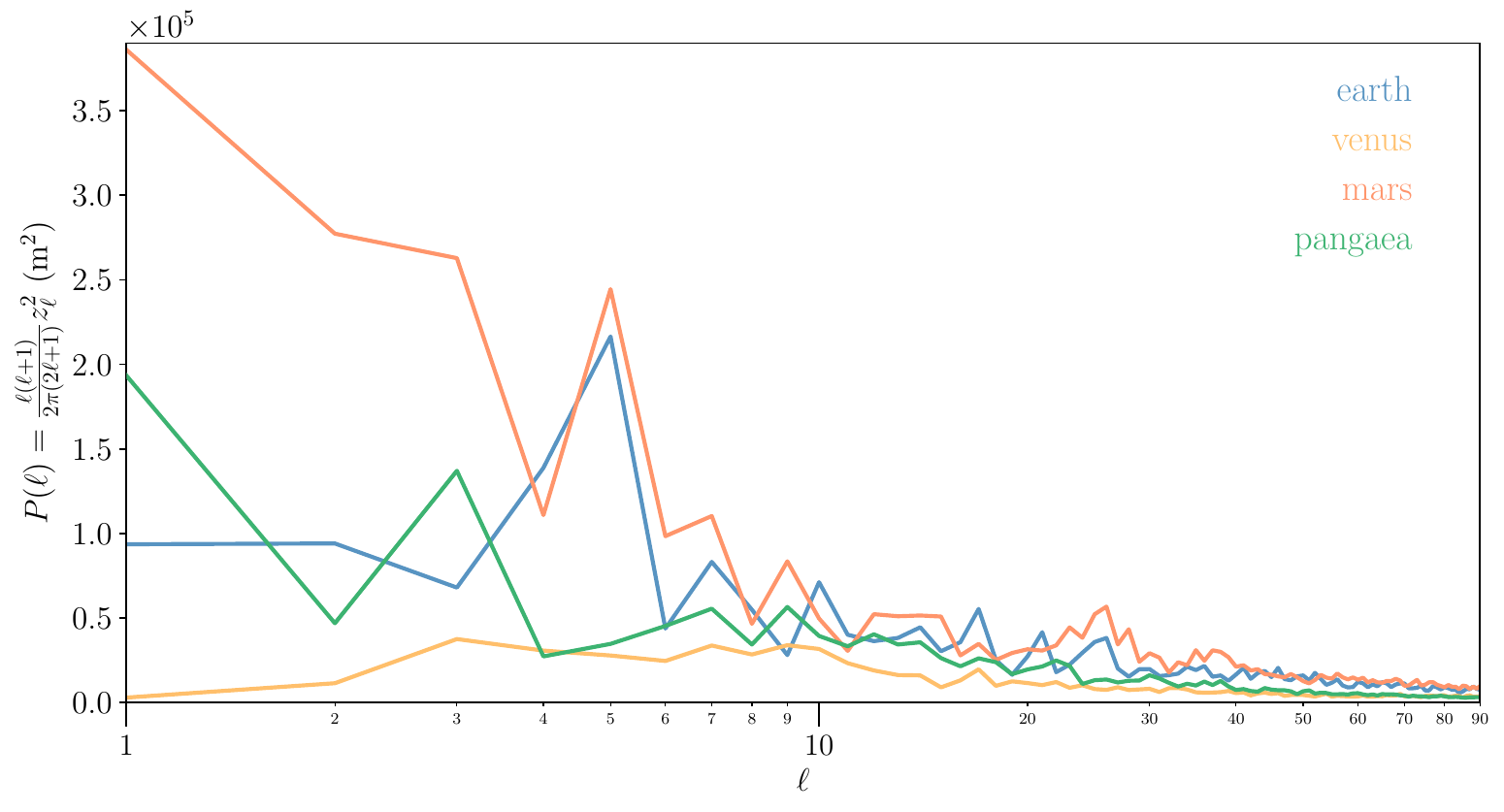}
\caption{The angular power spectra, as defined in Equation \ref{eq:power_spectrum}, of the elevation maps of present-day Earth in blue, Venus in yellow, and Mars in red. The topographic power spectrum of the Earth at 320 Mya, corresponding to the time of peak $\ell = 1$ spectral power, is shown in green.}
\label{fig:power_spectra}
\end{center}
\end{figure}

We represent the elevation data as a sum of real spherical harmonics:
\begin{equation}\label{eq:map_decomposition}
    z = \sum_{\ell, m} z\lm Y\lm\phimu
\end{equation}
where we adopt the geodesic normalization such that
\begin{equation}\label{eq:harmonic_normalization}
    \int\limits_{0}^{2\pi} \int\limits_{-1}^{1} Y\lm Y^{m'\star}_{\ell'}\,d\mu\,d\phi = 4 \pi \delta_{\ell \ell'} \delta_{m m'}.
\end{equation}
We also adopt the normalization convention of the angular power spectrum as
\begin{equation}\label{eq:power_spectrum}
    P\!\left(\ell\right) = \frac{\ell \left(\ell+1\right)}{2\pi \left(2\ell+1\right)} z^2_{\ell}.
\end{equation}
We plot the power spectra as defined in Equation \ref{eq:power_spectrum} for Earth, Venus, Mars, and Pangaea-era Earth in Figure \ref{fig:power_spectra}. These power spectra are kept non-normalized in order to show the true physical amplitudes of the elevation variance. We immediately see the relative flatness of Venus's surface, as well as the strong hemispheric contrast on Mars which leads to its dominant power at $\ell = 1$. Earth's highest power currently sits at $\ell = 5$, and highlights the contribution from the size and distribution of continents and ocean basins. Mars retains significant power out to $\ell = 5$ as well, a consequence of the distribution of its major volcanic provinces and extended low-elevation areas (notably the Tharsis Bulge and the Hellas basin). Further influence on the low-degree power comes from the shape of the boundary region between the high elevations in the south and the nearly uniform low-elevation region in the north.

\section{A Simple Model for Spectral Evolution after Tectonic Cessation}\label{sec:model} 
We develop a simple model to examine the broad effects on the topography, and its associated power spectrum, when tectonic activity ceases. The specific scenarios motivating this analysis are the forthcoming transition from the present oceanic climate to a runaway greenhouse climate on Earth \citep[see e.g.~the definitions and discussions in][]{kop13,kop14,Gillmann2014}, and the possible past occurrence of such an event on Venus. During these transitions, oceans suffer large-scale evaporation with ensuing photolysis of H$_2$O and loss of hydrogen to hydrodynamic escape. Venus could have experienced two runaway greenhouse periods during the first 50 Myr of the solar system \citep{Ramirez2016}, during which it could have lost to space a mass equivalent to $\sim\!3$--4 of Earth's oceans \citep{Ramirez2018}. Loss of water shuts down the tectonic cycle, the carbonate-silicate cycle ends, and the planet transitions to a stagnant-lid convection style \citep{Kasting1988a}.

Our model acts stochastically to infer the progression of structural changes that need to occur in a shift from an Earth-like to a Venus-like topographic spectrum, and we identify plausible physical processes that map to the stages and outcomes of our mathematical model. Modeling that employs random sampling has a history in describing the Venusian surface, a notable example being assessments of the distributions of craters. In works such as \citet{Strom1994,Romeo2013}, Monte Carlo simulations tested the hypothesis that crater locations on Venus are consistent with being drawn from a random spatial distribution. Our work has a parallel with this history, with our particular hypothesis being: the possible transition in Venus's surface, as described above, can be modeled in the power spectrum by processes that erode the features to some degree (attributed to a runaway greenhouse), then resurface with features that follow a size-frequency and mass-frequency distribution (attributed to flood-volcanic outflows in a post-active tectonic state).

The final configuration of our model is described below and represents the simplest parameterization that provides an accurate fit to the Venusian spectrum. Based on previous explorations, we ruled out a couple of model components that did not have significant constraints from the data:
\begin{itemize}
    \item There is no evidence to constrain any residual continent-ocean structure from Earth's original spectrum. In other words, our model fits just as well if all but the barest traces of the original spectrum are wiped out by the erosion component of our model.
    \item Accordingly, we find it is simplest to \emph{fix} the total strength of erosion in our model to that which preserves only the largest ($\ell=1$, dipole) moment of the spectrum. This allows us to estimate the timescale of erosion, as we discuss in \S \ref{sec:results:physical:erosion}.
    \item As a second corollary to the above, we find that the erosion and volcanism stages may be modeled as completely de-coupled periods. This does not imply that these processes did not co-occur, but rather that any volcanism occurring before or during the erosive period will not leave a significant impact on the modern topography.
    \item Finally, we find that there is no evidence for significant erosion in the post-greenhouse, volcanism-dominated era. We experimented with having two diffusion coefficients: one for the runaway greenhouse period, and the other for the volcanic period. The latter returned a very low value, consistent with negligible erosion; therefore, we remove it from our model.
\end{itemize}

\subsection{Model Parameterizations}\label{sec:model:parameters}
We consider two primary global-scale effects that would act during a stagnant-lid regime to reshape the topography: denudation of surface material with some associated time scale, and basaltic outflows from transient volcanic activity. The prescription is as follows:

\begin{enumerate}
    \item Over some time interval, generate some number $N$ of volcanic events. These events have masses, size scales, and positions on the globe, all drawn from their own probability distributions. For simplicity, we assume a constant mass density, such that all draws in mass are completely equivalent to draws in volume.
    \begin{enumerate}
        \item Outflow masses are drawn from a power-law distribution
            \begin{equation}\label{eq:mass_powerlaw}
                \frac{\mathop{dN}}{\mathop{d\bar{M}}} \propto \bar{M}^{-\nu} \,,
            \end{equation}
            where $\bar{M}$ is the ratio of the mass of the outflow province to some lower mass bound. One additional constraint is that the maximum mass is defined such that the expected number of ``volcanoes'' (a term that we use synonymously with ``volcanic provinces'') in the order of magnitude centered on our maximum mass is exactly unity. To implement this constraint in our modeling algorithm, we use $N$ to set the lower simulated bound on the mass ratio. This is not a \emph{physical} limit, but rather just the smallest mass of volcanic provinces that we choose to model, balancing computational demand with resolution of outflows needed to reproduce the spectrum to our limiting harmonic degree. The lower mass ratio limit is then given by
            \begin{equation}\label{eq:mass_lower-bound}
                \left(\frac{M_\mathrm{max}}{M_\mathrm{min}}\right)^{\nu-1} = 1 + N\left[10^{(\nu-1)/2} - 10^{-(\nu-1)/2}\right] \, ,
            \end{equation}
            which is valid for $\nu \neq 1$. To draw from this distribution given a random uniform sample, we use
            \begin{equation}\label{eq:mass_draw}
                \left\{ 1 - U\!\left(0, 1\right) \left[1 - \left(\frac{M_\mathrm{min}}{M_\mathrm{max}}\right)^{\nu-1}\right]\right\}^{-1/\left(\nu-1\right)}.
            \end{equation}
        \item For the outflow start times, the models have similar approaches. We normalize the volcanic interval to be some unit interval $\left[0, 1\right]$. Then, once the outflow masses are generated, each is simply assigned a uniform draw from the unit interval. This approximates a Poisson process where the time dependence with mass is accounted for in the choice of masses from the power-law distribution.
        \item For the shape and size of outflows, we start by assuming the outflow material has constant density $\rho$, such that the draws in masses are equivalent to draws in volume. All the outflows are ``Gaussian-like'' on the globe, in that we use the functional form of the symmetric Fisher-Bingham distribution, which is parametrized as
        \begin{equation}\label{eq:fisher-bingham}
            h_i\!\left(\phi,\theta,t\right) = \frac{M_i}{4\pi \rho R_\oplus^2} \frac{\kappa}{\sinh\kappa} e^{\kappa\left(\hat{r}\cdot\hat{r_i}\right)}.
        \end{equation}
        $h_i$ represents the vertical height of the volcano at each position, and $\kappa$ is a concentration parameter that is analogous to (though not exactly the same as) $1/\sigma^2$. The dot product $\hat{r}\cdot\hat{r_i}$ is equivalent to the cosine of the angle from the radial vector that points to the central peak of the volcano. One advantage of using this form for the volcano shape is that it is proportional to the generating function for the spherical harmonics; therefore, its transform is simply the conjugate of the harmonic functions $Y_{\ell}^{m}$ evaluated at the location of the volcano,
        \begin{equation}
            \left(h_i\right)\lm\!\left(t\right) =  \frac{I_{\ell+1/2}\!\left(\kappa\right)}{I_{1/2}\!\left(\kappa\right)} Y_{\ell}^{m\star}\!\left(\mu_i,\phi_i\right),
        \end{equation}
        where $I_n$ is the modified Bessel function of order $n$. We can alternatively express the concentration $\kappa$ in terms of the spherical harmonic degree scale, as
        \begin{equation}\label{eq:Lambda_parameter}
            \Lambda \equiv \pi \sqrt{\kappa}.
        \end{equation}
        Our samples for $\kappa$ depend on the corresponding samples in mass/volume. We start with some largest spatial scale, $\Lambda_\mathrm{max}$, that corresponds to the harmonic degree scale of the typical largest outflow, i.e.~that with mass $M_\mathrm{max}$. We then prescribe a power law with exponent $\lambda$ to connect with a mass scale:
            \begin{equation}\label{eq:size_powerlaw}
                \frac{\Lambda\!\left(M\right)}{\Lambda_\mathrm{max}} = \left(\frac{M}{M_\mathrm{max}}\right)^{-\lambda}.
            \end{equation}
            In this model, $\lambda = 1/3$ will produce outflows that are self-similar in shape as mass is varied. An exponent $\lambda < 1/3$ means that smaller outflows tend to be flatter (more platykurtic) in profile than larger ones, and accordingly $\lambda > 1/3$ would mean the smaller the outflow, the pointier (more leptokurtic) the outflow will be. Here we also do not explicitly model the time dependence of the outflows. 
        \item We use a uniform distribution on the surface of a sphere for generating positions in co-latitude $\theta$ and longitude $\phi$
        \begin{gather}
        \begin{aligned}\label{eq:position_distribution}
            \theta &\leftarrow \cos^{-1}\!\left[U\!\left(1, -1\right)\right] \\
                   &= \cos^{-1}\!\left[1 - 2 U\!\left(0, 1\right)\right], \\
            \phi   &\leftarrow U\!\left(0, 2\pi\right) \\
                   &= 2\pi U\!\left(0, 1\right).
        \end{aligned}
        \end{gather}
    \end{enumerate}
    \item Over longer timescales, evolve the topography via diffusion on a sphere via
    \begin{equation}
        \frac{\partial z}{\partial t} = \omega_\mathrm{dif} \nabla_{\mathrm{H}}^2 z
    \end{equation}
    where $\omega_\mathrm{dif} \equiv D/R_\oplus^2$ for a diffusion coefficient $D$, which is assumed to be a constant, and with the horizontal Laplacian operator given by
    \begin{equation}
        \nabla^2_{\mathrm{H}} \equiv \frac{1}{R_\oplus^2 \sin^2\theta} \left[ \sin\theta \frac{\partial}{\partial \theta}\left(\sin\theta \frac{\partial}{\partial \theta}\right) + \frac{\partial^2}{\partial \phi^2}\right].
    \end{equation}
    It is easiest to consider the effects of the diffusion on the coefficients, since spherical harmonics are eigenfunctions of the Laplacian:
    \begin{equation}
        z\lm\!\left(t\right) = z\lm\!\left(t_0\right) d_{\ell}\!\left(t-t_0\right)
    \end{equation}
    where 
    \begin{equation}
        d_{\ell}\!\left(t\right) \equiv \exp\!\left[-\ell\left(\ell+1\right)\omega_\mathrm{dif} t\right].
    \end{equation}
    The full analytical solution to the map coefficients for a simulation differs slightly between the models. We fix our diffusion coefficient to the value that reduces the $\ell=1$ amplitude in Earth's power spectrum to that of Venus --- one for the initial erosion interval, and one that occurs during the subsequent outflow period. This leads to a reduction in the coefficients at the time at which erosion ceases (call it $t_\mathrm{eros}$):
        \begin{equation}
            z\lm\!\left(t_\mathrm{eros}\right) = z\lm\!\left(0\right) d_{\ell}\!\left(\omega t = \Delta\right).
        \end{equation}
    The value that produces a Venus-like $\ell=1$ power corresponds to $\Delta \approx 1.86$.
\end{enumerate}
The model thus has 5 free parameters: (1) $N$, the number of events; (2) $\log_{10}\!\left(V_\mathrm{max}/V_\oplus\right)$, which yields the mass of the largest event $M_\mathrm{max}$; (3) $\nu$, the power-law exponent of how the masses of events scale with their frequencies, as described in Equations \ref{eq:mass_powerlaw}--\ref{eq:mass_draw}; (4) $\Lambda_\mathrm{max}$, the characteristic spherical harmonic degree of the largest outflow; and (5) $\lambda$, the power-law exponent of how the harmonic degree (related to the concentration parameter via Equation \ref{eq:Lambda_parameter}) scales with mass/volume, as described in Equation \ref{eq:size_powerlaw}.

\subsection{Model Fitting}\label{sec:model:fitting}
We perform our parameter estimation using a nested sampling algorithm, using the \texttt{dynesty} Python package \citep{Speagle2020}. Since the model is stochastic, we set a fixed seed for each uniform draw, and generate a composite spectrum from the mean and standard deviation of 10 successive draws with some set of parameters. The uncertainties in the fitting were determined by adding the variances from both Venus's power spectrum, as well as the stochastic variance from the models at each harmonic degree. We estimate the uncertainties on Venus's power spectrum via the Magellan data \citep{VenusDEM} by drawing 1000 realizations of the spectrum from a normal distribution centered on the mean spectrum and with the variance given by the errors. The resulting uncertainty estimates follow a linear trend in degree, with a fractional uncertainty of approximately $\left(1823\pm17\right) + \left(861\pm2\right)\ell$ parts per million.

We run the nested sampler until the default stopping criteria are satisfied, which depends on the heuristic relative importance of estimating the posterior distributions versus the evidence, and how many effective samples are desired. More details on the stopping function can be found in the \href{https://dynesty.readthedocs.io/en/stable/dynamic.html#stopping-function}{Dynesty documentation regarding stopping functions}. Each run was initialized with 80 live points, and allowed to run with an effectively unlimited ($5\times10^5$) number of effective samples before stopping. In practice, the number of effective samples, as determined by those that contain at least 95\% of the importance in the samples, is $\sim$500.

\section{Results}\label{sec:results}
\subsection{Constraints on the Power Spectrum Evolution}\label{sec:results:dimensionless}
\begin{figure}
\begin{center}
\includegraphics[width=17cm]{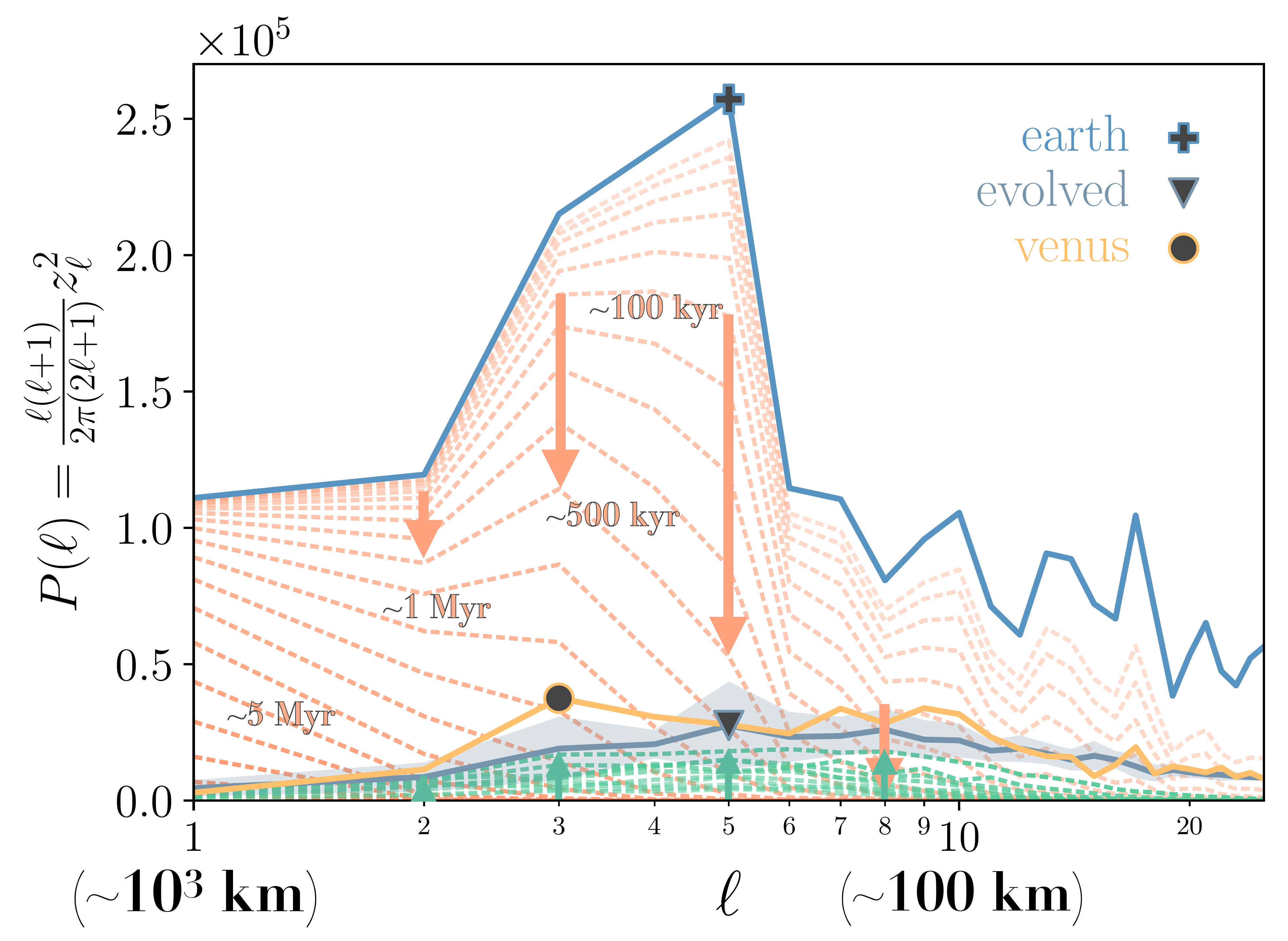}
\caption{The power spectra of the uniform weighting map shown in the central row of Figure \ref{fig:model_maps}. The shaded regions around the evolved Earth spectra represent the 1$\sigma$ ranges of a set of 100 randomized simulations. The dashed lines highlight that, under the constraints of this model and its assumptions, the transformation of the surface consists of two essentially distinct steps: the smoothing of the continent/ocean bimodality via diffusion (shown by the red dashed lines), and the subsequent build-up of surface features through volcanic outflows (in green). This result implies that, while our model nominally assumes a constant surface diffusion rate and a fixed period of volcanism, the results will be virtually identical for equal values of the total diffusion (parametrized by $\Delta$ in Table \ref{table:model_parameters}) and total volcanic outflow mass.}
\label{fig:model_spectra}
\end{center}
\end{figure}

\begin{figure}
\begin{center}
\includegraphics[width=17cm]{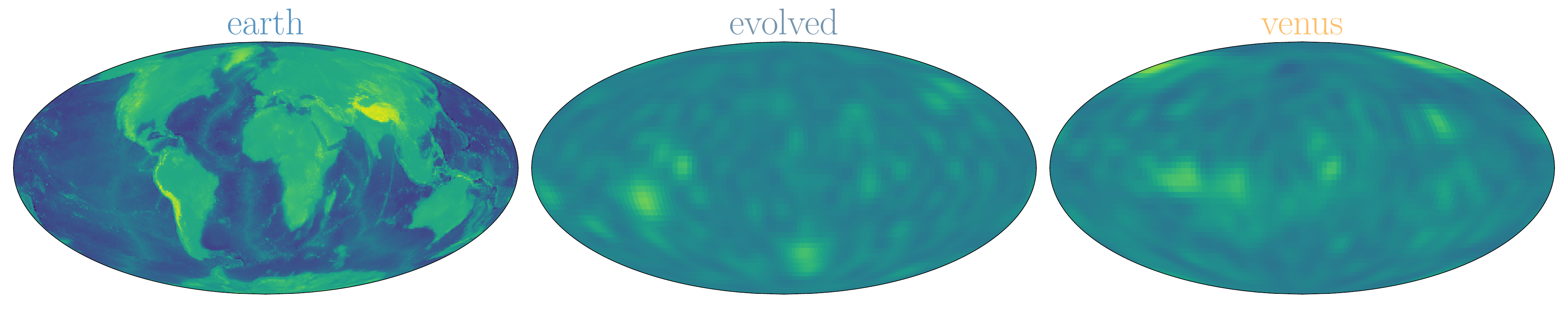}
\caption{A global map from the stochastic model. The left map is color-coded by Earth's current topography, the right map by Venus's current topography (using the same color scale and spatial resolution as the model map), and the central map show Earth's topography evolved using our routine. The evolved map shown here is the model with the smallest residuals of the runs used to generate the mean spectrum for the maximum likelihood parameters.}
\label{fig:model_maps}
\end{center}
\end{figure}

The model captures the shape of the spectrum out to our maximum fitting degree $\ell = 25$, though fails to capture the strength of the $\ell = 3$ mode in particular (Figure \ref{fig:model_spectra}). This points to the inability of our model to reproduce the structure of the terrae, the largest Venusian features. The terrae are extended and irregularly shaped and therefore it may be no surprise that our model cannot capture their influence on the spectrum. Looking at the two maps side-by-side at this spatial resolution (Figure \ref{fig:model_maps}), we see that the terrae show up as clustered highland regions. Beyond this, however, and because volcanic features are thought to shape the vast majority of the remainder of Venus's surface, our model produces a spectrum that matches Venus's out to our limiting size scale.

\begin{figure}
\begin{center}
\includegraphics[width=17cm]{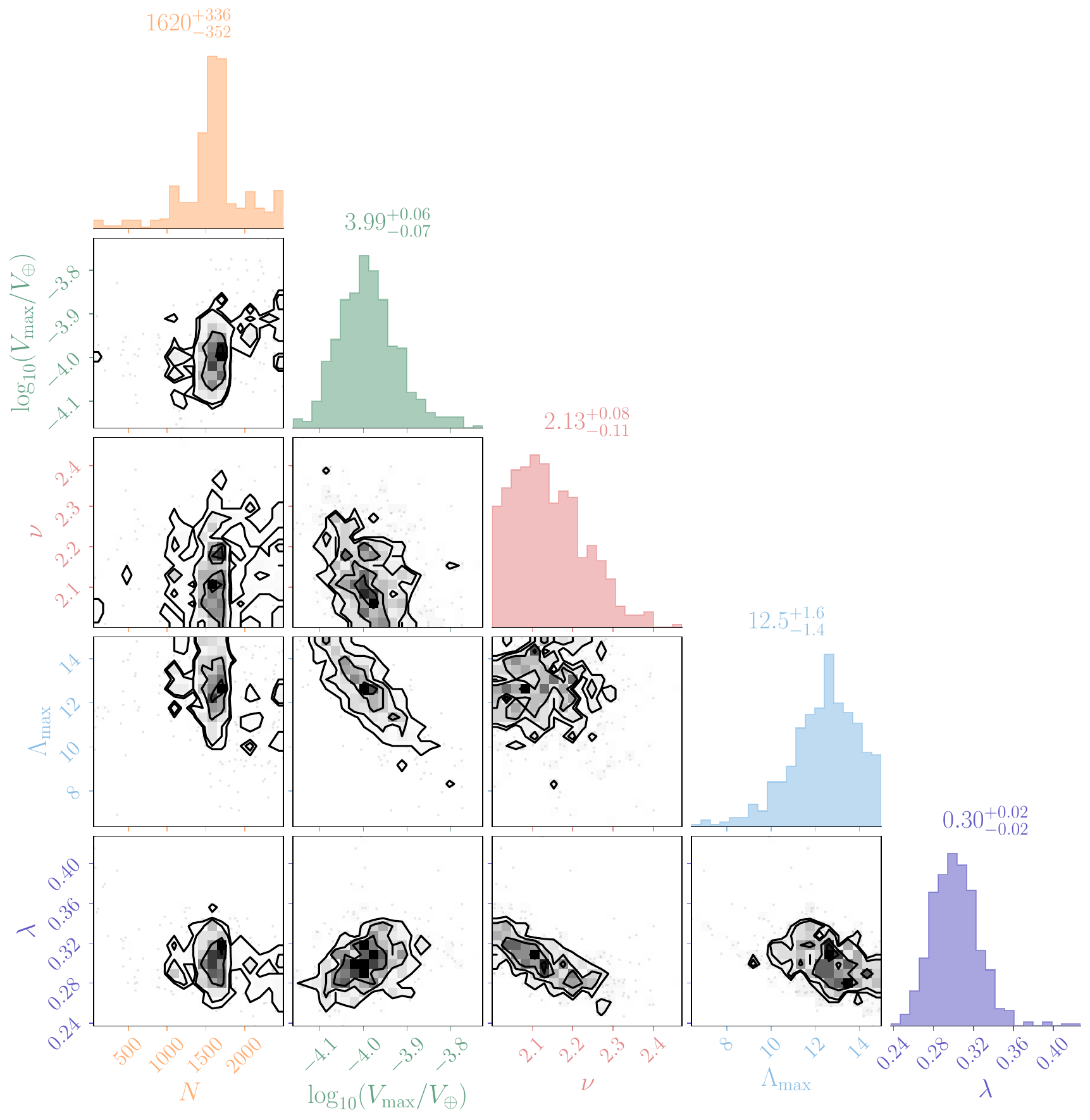}
\caption{A corner plot of the retrieved posterior distributions for the dimensionless parameters (as defined in \S \ref{sec:model:parameters}) of the diffusion+volcanic outflow model. The median and 68\% confidence interval are shown above each histogram.}
\label{fig:model_corners}
\end{center}
\end{figure}

The constraints from the nested sampling parameter estimation are shown in Table \ref{table:model_parameters} and Figure \ref{fig:model_corners}. The number of events is the least well constrained, as measured by fractional uncertainty in the posterior distributions. $N$ however does not show any strong correlation with the remaining parameters, though there is a hint of a positive correlation between $N$ and the volume of the largest outflow ($V_\mathrm{max}$). This correlation suggests the model finds a preferred minimum outflow scale, and increases or decreases the number of events as the largest outflow volume grows or shrinks. Negative correlations appear between both the maximum volume and size scales, as well as between the power-law exponents for volume and size. The first is not surprising, as we would expect larger outflows in volume may also tend to be larger in area. The anti-correlation between the power-law exponents suggests that as the outflows are weighted more toward lower-mass events, the model compensates by flattening the outflows so that they maintain nearly the same area. However, it should be noted that the 68\% confidence intervals of each of these parameters is of order a few percent; the absolute values of the parameters thus appear to be fairly well constrained, despite the small number of realizations used in the fitting routine. The following sub-section details how the constraints on these model parameters, along with the amount of diffusion estimated from the $\ell=1$ modes, can be translated into an estimate for physical timescales and outflow properties.

\begin{table}\normalsize
\centering
\begin{tabular}{lr|lr}
\hline
\multicolumn{2}{c}{\textbf{Dimensionless}} & \multicolumn{2}{c}{\textbf{Physical}} \\
\multicolumn{1}{l}{Name} & \multicolumn{1}{r}{Value} & \multicolumn{1}{l}{Name} & \multicolumn{1}{r}{Value} \\
\hline
$N$ & $1616^{+820}_{-1191}$ & \\
$\log_{10}\!V$ & $-3.99^{+0.17}_{-0.12}$ & $V_\mathrm{max}$ & $1.09^{+0.67}_{-0.37} \times 10^{8}$ km$^3$ \\
$\nu$ & $2.68^{+0.65}_{-0.31}$ & $V_\mathrm{tot}$ & $5.44^{+8.86}_{-1.93} \times 10^{8}$ km$^3$ \\
$\Lambda_\mathrm{max}$ & $7.95^{+8.56}_{-5.03}$ & $t_\mathrm{out}$ & $2.72^{+4.43}_{-0.97} \times 10^{8}$ yr \\
$\lambda$ & $0.61^{+0.25}_{-0.16}$ & \\
$\Delta$ & 1.83 (fixed) & $t_\mathrm{eros}$ & $5.1^{+1.8}_{-1.1} \times 10^{6}$ yr \\
\hline
\end{tabular}
\caption{Dimensionless parameters used in the Earth evolution model, along with the corresponding physical volume parameters. $V_\mathrm{max}$ is the volume of the most massive outflow expected to occur exactly once during the volcanic interval; $V_\mathrm{tot}$ is the total integrated volume across all modeled spatial scales with the retrieved mass scale and expected number of events.}
\label{table:model_parameters}
\end{table}

\subsection{Estimating Timescales of Possible Earth-to-Venus Transitions}\label{sec:results:physical} 
\subsubsection{The Erosion Timescale}\label{sec:results:physical:erosion}
Modern Earth has an active hydrological cycle, and accordingly we can make order-of-magnitude estimates for resurfacing due to sediment deposition using characteristic basin sizes and sediment loads for several of Earth's major rivers. We estimate that, if this process were global and acted alone to diffuse Earth's topography, Earth would attain the muted surface variation of Venus in a geologically relatively brief time scale of order $\sim\!5$ Myr. This estimate relies on estimates of global denudation on modern Earth. In \citet{Harel2016}, \S 3.1, we find that the global data returns a mean of $242\pm59$ m Myr$^{-1}$. Translating the total variance in the first 25 degrees of each spectra to a characteristic relief scale ($\sqrt{\sigma_z^2} \approx 1558$ m for Earth, 655 m for Venus), we find an erosion time of $5.1^{+1.8}_{-1.1}$ Myr. However, this relies heavily on the assumption that Venus's greenhouse state would have comparable erosion rates to those of the basins of Earth's bedrock rivers; we discuss further in \S \ref{sec:discussion}. 

\subsubsection{The Erosion Timescale}\label{sec:results:physical:volcanism}
The largest outflow expected to occur exactly once across the volcanic interval is $\sim 0.01\%$ the total volume of the planet, with a retrieved distribution of $1.09^{+0.67}_{-0.37} \times 10^{8}$ km$^3$ and a total modeled outflow of $5.44^{+8.86}_{-1.93} \times 10^{8}$ km$^3$ (see Table \ref{table:model_parameters}, right column). In order to obtain our estimate of how long our model predicts volcanism has taken place since the catastrophic resurfacing, we need to match our outflow volumes with an estimate of the mean outflow rate. \citet{Deligne2015} put Earth's current global outflow rates at $\sim\!20$ km$^3$ yr$^{-1}$ at the mid-oceanic ridges, $\sim\!2$ km$^3$ yr$^{-1}$ from volcanic hotspots, and $\sim\!0.5$ km$^3$ yr$^{-1}$ at subduction zones. If we assume a mean outflow rate of $2$ km$^3$ yr$^{-1}$ to match modern Earth's volcanic hotspots, we estimate that volcanism has occurred for the past $272^{+443}_{-97}$ Myr. However, these rates are high compared with the estimates from analyses of the effect of volcanic resurfacing on crater size distributions, which yield rates of order $\sim$0.01--1 km$^3$/year \citep{Phillips1992,Bullock1993,Strom1994,Basilevsky1997}. At the high end of this range, our inferred duration of volcanism doubles to $544^{+886}_{-193}$ Myr, but is still consistent with the 300--600 Mya range quoted in works such as \citet{Nimmo1998}. However, even assuming the high end of a more conservative range such as that in \citet{Strom1994}, at 0.01--0.15 km$^3$ yr$^{-1}$, our median model estimate would require a period of volcanism approaching the age of the Solar System.


\section{Discussion}\label{sec:discussion}
\subsection{The Erosion Timescale in Context}\label{sec:discussion:erosion}
As noted above, our result of an erosion timescale $\sim$ 5 Myr relies on assuming that rates of denudation in bedrock rivers on modern Earth are a valid proxy for a sustained planet-wide erosion due to a runaway greenhouse state. In contrast, on a dry, stagnant-lid planet such as modern-day Venus that has low surface wind speeds and minimal diurnal temperature variations, erosion is inefficient to the point of being negligible \citep{Sagan1976, Strom1993}. Our work relies on the assumption there was such a period as a runaway greenhouse catastrophe where fluvial erosion greatly outstripped tectonic uplift and volcanic emplacement. Theoretical studies of runaway greenhouse scenarios find that during the early phases of such a scenario, erosion rates are elevated by the increase in temperature and the increase in overall precipitation \citep{nearing2004expected}. Calculations of hydrodynamic escape rates for a potential runaway greenhouse on early Venus yield a time scale similar to what our model returns: $\sim\!10$ Myr for a water mass equivalent to Earth's current oceans \citep[see for example][]{Ribas2005, Ramirez2018}. Therefore, under at least simple estimates of runaway greenhouse water loss, one could expect that the catastrophic transition associated with evaporation of the oceans could provide sufficient conditions for continental erosion.

\subsection{The Volcanic Outflow Timescale in Context}\label{sec:discussion:volcanism}
Just as the history of Venus's tectonics are still an active subject of debate, so too is the history of volcanism. While we can use our model to estimate a cumulative amount of resurfacing from volcanic outflows, and can put a constraint on the frequency as a function of mass/volume, it does not account for any variation in the outflow rate. The climate model analyses of \citet{Way2020} and \citet{Way2022} suggest that the resurfacing of Venus was immediately preceded by the onset of multiple large igneous provinces (LIPs), with this LIP period itself lasting of order tens to millions of years. Despite these inferences, we have to date never observed evidence of an active volcanic eruption on the surface of Venus, and estimating recent eruption frequencies relies on assumptions about the tectonic history as well as extrapolation from Earth's volcanism \citep[see e.g.][]{Byrne2022,King2022}. The VERITAS and DAVINCI missions \citep{Smrekar2020,Garvin2022} are planned missions for a Venus orbiter and in-situ probe, respectively, that will allow us to observe the effects of active volcanism on Venus should it be present. For now, we have a range of estimates of the global outflow rates from cratering studies spanning two orders of magnitude, which introduces a high degree of uncertainty to our timescales.

A parallel constraint on the rate of volcanism comes from the atmospheric composition, with the present atmospheric concentration of sulfur compounds (SO$_2$, and the H$_2$SO$_4$ present in the global cloud layer) thought to depend on a continual outgassing from volcanoes \citep{Fegley1989,King2022}. Most recently, \citet{Gillmann2022} and \citet{Weller2023} have proposed that the current abundance of N$_2$ and CO$_2$ in Venus's modern atmosphere strongly implies the historical existence of more active tectonics that persisted for at least $\sim$1 Gyr that allowed for a much higher outgassing rate than is inferred at present day. As with the aforementioned future probes of Venus's current volcanic activity, one can imagine that a mission like DAVINCI will be invaluable for directly measuring atmospheric composition, and therefore further constraining the possible histories Venus could have undergone.

\section{Conclusion}\label{sec:conclusion}
We present the results of a model that evolves the power spectrum of a terrestrial planet's topography according to a reproduction of global-scale changes that would occur during and after a runaway greenhouse state. By applying this to Earth's modern power spectrum we evaluate the plausibility that, by resurfacing of an Earth-like surface by a combination erosion and volcanism, the cessation of plate tectonics could lead to a Venusian landscape. We find that:
\begin{itemize}
\item By treating global erosion with a simple diffusion model, and tuning its result to current Earth's mean denudation rates for bedrock river basins, we retrieve a timescale of $5.1^{+1.8}_{-1.1}$ Myr to erode Earth's relief structure into one resembling Venus's. \item By fitting our volcanic outflow model, we estimate an outflow timescale of $544^{+886}_{-193}$ Mya, assuming a mean outflow rate of 1 km$^3$ yr$^{-1}$. This is consistent with the scale of times predicted in the literature.
\item Given the strength of the erosion in our model, the model is largely insensitive to the initial continental structure, and the amount of erosion following the greenhouse state is also not well constrained.
\end{itemize}
Given the ongoing research efforts on Venus's tectonic, volcanic, and atmospheric histories, as discussed in \S \ref{sec:discussion}, we expect the approach of analyzing and modeling the effects of resurfacing on topographic power spectra to provide a statistical complement to the focused interior, surface, and climate modeling efforts that probe the precise outcomes of physical processes on Venus. One could imagine constructing a model that can stochastically simulate a sample of terrestrial surfaces throughout multiple stages of evolution, for predicting the outcomes of observables for terrestrial-sized exoplanets. Such a model would be readily adapted to new insights gained from the ongoing research into the evolution of the terrestrial planets in our own Solar System.

\begin{acknowledgments}
GL acknowledges generous support from the Heising-Simons Foundation through Grant 2021-2802 to Yale University.
\end{acknowledgments}

\vspace{5mm}

\software{\texttt{astropy} \citep{ast13}, \texttt{cartopy} \citep{Cartopy}, \texttt{dynesty} \citep{Speagle2020}, \texttt{numpy} \citep{van11}, \texttt{scipy} \citep{jon01,mck10}, \texttt{SHTools} \citep{Wieczorek2018}}


\bibliography{library}{}

\begin{thebibliography}{}
\expandafter\ifx\csname natexlab\endcsname\relax\def\natexlab#1{#1}\fi
\providecommand{\url}[1]{\href{#1}{#1}}
\providecommand{\dodoi}[1]{doi:~\href{http://doi.org/#1}{\nolinkurl{#1}}}
\providecommand{\doeprint}[1]{\href{http://ascl.net/#1}{\nolinkurl{http://ascl.net/#1}}}
\providecommand{\doarXiv}[1]{\href{https://arxiv.org/abs/#1}{\nolinkurl{https://arxiv.org/abs/#1}}}

\bibitem[{Amante \& Eakins(2009)}]{Amante2009}
Amante, C., \& Eakins, B.~W. 2009, {ETOPO1 1 Arc-Minute Global Relief Model:
  procedures, data sources and analysis}, Tech. rep., National Geophysical Data
  Center, Marine Geology and Geophysics Division, Boulder, Colorado,
  \dodoi{10.7289/V5C8276M}

\bibitem[{{Astropy Collaboration} {et~al.}(2013){Astropy Collaboration},
  Robitaille, Tollerud, Greenfield, Droettboom, Bray, Aldcroft, Davis,
  Ginsburg, Price-Whelan, Kerzendorf, Conley, Crighton, Barbary, Muna,
  Ferguson, Grollier, Parikh, Nair, Unther, Deil, Woillez, Conseil, Kramer,
  Turner, Singer, Fox, Weaver, Zabalza, Edwards, {Azalee Bostroem}, Burke,
  Casey, Crawford, Dencheva, Ely, Jenness, Labrie, Lim, Pierfederici, Pontzen,
  Ptak, Refsdal, Servillat, \& Streicher}]{ast13}
{Astropy Collaboration}, Robitaille, T., Tollerud, E., {et~al.} 2013, \aap,
  558, A33, \dodoi{10.1051/0004-6361/201322068}

\bibitem[{Balmino(1993)}]{Balmino1993}
Balmino, G. 1993, Geophys. Res. Lett., 20, 1063, \dodoi{10.1029/93GL01214}

\bibitem[{Balmino {et~al.}(1973)Balmino, Lambeck, \& Kaula}]{Balmino1973}
Balmino, G., Lambeck, K., \& Kaula, W.~M. 1973, J. Geophys. Res., 78, 478,
  \dodoi{10.1029/JB078i002p00478}

\bibitem[{Basilevsky \& Head(1998)}]{Basilevsky1998}
Basilevsky, A.~T., \& Head, J.~W. 1998, J. Geophys. Res. Planets, 103, 8531,
  \dodoi{10.1029/98JE00487}

\bibitem[{Basilevsky {et~al.}(1997)Basilevsky, Head, Schaber, \&
  Strom}]{Basilevsky1997}
Basilevsky, A.~T., Head, J.~W., Schaber, G.~G., \& Strom, R.~G. 1997, in Venus
  II, ed. S.~W. Bougher, D.~M. Hunten, R.~J. Phillips, M.~S. Matthews, A.~S.
  Ruskin, \& M.~L. Guerrieri (University of Arizona Press), 1047--1084,
  \dodoi{10.2307/j.ctv27tct5m.35}

\bibitem[{Bercovici(2003)}]{Bercovici2003}
Bercovici, D. 2003, Earth Planet. Sci. Lett., 205, 107,
  \dodoi{10.1016/S0012-821X(02)01009-9}

\bibitem[{Bercovici {et~al.}(2015)Bercovici, Tackley, \&
  Ricard}]{Bercovici2015b}
Bercovici, D., Tackley, P., \& Ricard, Y. 2015, in Treatise Geophys., Vol.~7
  (Elsevier), 271--318, \dodoi{10.1016/B978-0-444-53802-4.00135-4}

\bibitem[{Bills \& Kiefer(1985)}]{Bills1985b}
Bills, B., \& Kiefer, W. 1985, in Lunar Planet. Sci. Conf., 59--60

\bibitem[{Bills \& Ferrari(1978)}]{Bills1978}
Bills, B.~G., \& Ferrari, A.~J. 1978, J. Geophys. Res. Solid Earth, 83, 3497,
  \dodoi{10.1029/JB083iB07p03497}

\bibitem[{Bills {et~al.}(1993)Bills, Frey, Kiefer, Nerem, \& Zuber}]{Bills1993}
Bills, B.~G., Frey, H.~V., Kiefer, W.~S., Nerem, R.~S., \& Zuber, M.~T. 1993,
  in Lunar Planet. Sci. Conf., 105

\bibitem[{Bills \& Kobrick(1985)}]{Bills1985a}
Bills, B.~G., \& Kobrick, M. 1985, J. Geophys. Res. Solid Earth, 90, 827,
  \dodoi{10.1029/JB090iB01p00827}

\bibitem[{Bills \& Lemoine(1995)}]{Bills1995a}
Bills, B.~G., \& Lemoine, F.~G. 1995, J. Geophys. Res., 100, 26275,
  \dodoi{10.1029/95JE02982}

\bibitem[{Bills \& Nerem(1995)}]{Bills1995b}
Bills, B.~G., \& Nerem, R.~S. 1995, J. Geophys. Res., 100, 26317,
  \dodoi{10.1029/95JE03101}

\bibitem[{Bjonnes {et~al.}(2012)Bjonnes, Hansen, James, \&
  Swenson}]{Bjonnes2012}
Bjonnes, E., Hansen, V., James, B., \& Swenson, J. 2012, Icarus, 217, 451,
  \dodoi{10.1016/j.icarus.2011.03.033}

\bibitem[{Bottke {et~al.}(2015)Bottke, Ghent, Mazrouei, Robbins, \&
  Vokrouhlicky}]{Bottke2015}
Bottke, W., Ghent, R., Mazrouei, S., Robbins, S., \& Vokrouhlicky, D. 2015, in
  AAS/Division Planet. Sci. Meet. Abstr. \#47, AAS/Division for Planetary
  Sciences Meeting Abstracts, 201.07

\bibitem[{Buck(2015)}]{Buck2015}
Buck, W. 2015, in Treatise Geophys., Vol.~6 (Elsevier), 325--379,
  \dodoi{10.1016/B978-0-444-53802-4.00118-4}

\bibitem[{Bullock {et~al.}(1993)Bullock, Grinspoon, \& Head}]{Bullock1993}
Bullock, M.~A., Grinspoon, D.~H., \& Head, J.~W. 1993, Geophys. Res. Lett., 20,
  2147, \dodoi{10.1029/93GL02505}

\bibitem[{Byrne {et~al.}(2021)Byrne, Ghail, Şeng{\"{o}}r, James, Klimczak, \&
  Solomon}]{Byrne2021}
Byrne, P.~K., Ghail, R.~C., Şeng{\"{o}}r, A. M.~C., {et~al.} 2021, Proc. Natl.
  Acad. Sci., 118, e2025919118, \dodoi{10.1073/pnas.2025919118}

\bibitem[{Byrne \& Krishnamoorthy(2022)}]{Byrne2022}
Byrne, P.~K., \& Krishnamoorthy, S. 2022, J. Geophys. Res. Planets, 127, 1,
  \dodoi{10.1029/2021JE007040}

\bibitem[{Condie \& Kr{\"{o}}ner(2008)}]{Condie2008}
Condie, K.~C., \& Kr{\"{o}}ner, A. 2008, Spec. Pap. 440 When Did Plate
  Tectonics Begin Planet Earth?, 2440, 281, \dodoi{10.1130/2008.2440(14)}

\bibitem[{Deligne \& Sigurdsson(2015)}]{Deligne2015}
Deligne, N.~I., \& Sigurdsson, H. 2015, in Encycl. Volcanoes, second edi edn.
  (Elsevier), 265--272, \dodoi{10.1016/B978-0-12-385938-9.00014-6}

\bibitem[{Dietz(1961)}]{Dietz1961}
Dietz, R.~S. 1961, Nature, 190, 854, \dodoi{10.1038/190854a0}

\bibitem[{Driscoll \& Bercovici(2013)}]{Driscoll2013}
Driscoll, P., \& Bercovici, D. 2013, Icarus, 226, 1447,
  \dodoi{10.1016/j.icarus.2013.07.025}

\bibitem[{Fegley \& Prinn(1989)}]{Fegley1989}
Fegley, B., \& Prinn, R.~G. 1989, Nature, 337, 55, \dodoi{10.1038/337055a0}

\bibitem[{Foley \& Driscoll(2016)}]{Foley2016}
Foley, B.~J., \& Driscoll, P.~E. 2016, Geochemistry, Geophys. Geosystems, 17,
  1885, \dodoi{10.1002/2015GC006210}

\bibitem[{Foley \& Smye(2018)}]{Foley2018}
Foley, B.~J., \& Smye, A.~J. 2018, Astrobiology, 18, 873,
  \dodoi{10.1089/ast.2017.1695}

\bibitem[{Gagnon {et~al.}(2006)Gagnon, Lovejoy, \& Schertzer}]{Gagnon2006}
Gagnon, J.-S., Lovejoy, S., \& Schertzer, D. 2006, Nonlinear Process. Geophys.,
  13, 541, \dodoi{10.5194/npg-13-541-2006}

\bibitem[{Garvin {et~al.}(2022)Garvin, Getty, Arney, Johnson, Kohler, Schwer,
  Sekerak, Bartels, Saylor, Elliott, Goodloe, Garrison, Cottini, Izenberg,
  Lorenz, Malespin, Ravine, Webster, Atkinson, Aslam, Atreya, Bos,
  Brinckerhoff, Campbell, Crisp, Filiberto, Forget, Gilmore, Gorius, Grinspoon,
  Hofmann, Kane, Kiefer, Lebonnois, Mahaffy, Pavlov, Trainer, Zahnle, \&
  Zolotov}]{Garvin2022}
Garvin, J.~B., Getty, S.~A., Arney, G.~N., {et~al.} 2022, Planet. Sci. J., 3,
  117, \dodoi{10.3847/PSJ/ac63c2}

\bibitem[{Gillmann \& Tackley(2014)}]{Gillmann2014}
Gillmann, C., \& Tackley, P. 2014, J. Geophys. Res. E Planets, 119, 1189,
  \dodoi{10.1002/2013JE004505}

\bibitem[{Gillmann {et~al.}(2022)Gillmann, Way, Avice, Breuer, Golabek,
  H{\"{o}}ning, Krissansen-Totton, Lammer, O'Rourke, Persson, Plesa, Salvador,
  Scherf, \& Zolotov}]{Gillmann2022}
Gillmann, C., Way, M.~J., Avice, G., {et~al.} 2022, Space Sci. Rev., 218, 56,
  \dodoi{10.1007/s11214-022-00924-0}

\bibitem[{G{\'{o}}rski {et~al.}(2018)G{\'{o}}rski, Bills, \&
  Konopliv}]{Gorski2018}
G{\'{o}}rski, K.~M., Bills, B.~G., \& Konopliv, A.~S. 2018, Planet. Space Sci.,
  160, 84, \dodoi{10.1016/j.pss.2018.03.015}

\bibitem[{Harel {et~al.}(2016)Harel, Mudd, \& Attal}]{Harel2016}
Harel, M.-A., Mudd, S., \& Attal, M. 2016, Geomorphology, 268, 184,
  \dodoi{10.1016/j.geomorph.2016.05.035}

\bibitem[{Harrison {et~al.}(1981)Harrison, Brass, Saltzman, Sloan, Southam, \&
  Whitman}]{Harrison1981}
Harrison, C., Brass, G., Saltzman, E., {et~al.} 1981, Earth Planet. Sci. Lett.,
  54, 1, \dodoi{10.1016/0012-821X(81)90064-9}

\bibitem[{Hauck {et~al.}(1998)Hauck, Phillips, \& Price}]{Hauck1998}
Hauck, S.~A., Phillips, R.~J., \& Price, M.~H. 1998, J. Geophys. Res. Planets,
  103, 13635, \dodoi{10.1029/98JE00400}

\bibitem[{Head {et~al.}(1981)Head, Yuter, \& Solomon}]{Head1981}
Head, J., Yuter, S., \& Solomon, S. 1981, Am. Sci., 69, 614

\bibitem[{{Head III}(1994)}]{Head1994}
{Head III}, J.~W. 1994, Nature, 372, 729, \dodoi{10.1038/372729a0}

\bibitem[{Hess(1962)}]{Hess1962}
Hess, H.~H. 1962, Petrol. Stud.
\newblock \url{https://ci.nii.ac.jp/naid/10014849678/en/}

\bibitem[{Hirt \& Rexer(2015)}]{Hirt2015}
Hirt, C., \& Rexer, M. 2015, Int. J. Appl. Earth Obs. Geoinf., 39, 103,
  \dodoi{10.1016/j.jag.2015.03.001}

\bibitem[{Jones {et~al.}(2001)Jones, Oliphant, Peterson, \& Others}]{jon01}
Jones, E., Oliphant, T., Peterson, P., \& Others. 2001, {{SciPy}: Open source
  scientific tools for {Python}}.
\newblock \url{http://www.scipy.org/}

\bibitem[{Kasting(1988)}]{Kasting1988a}
Kasting, J.~F. 1988, Icarus, 74, 472, \dodoi{10.1016/0019-1035(88)90116-9}

\bibitem[{King(2018)}]{King2018}
King, S.~D. 2018, J. Geophys. Res. Planets, 123, 1041,
  \dodoi{10.1002/2017JE005475}

\bibitem[{King(2022)}]{King2022}
---. 2022, J. Geophys. Res. Planets, 127, \dodoi{10.1029/2022JE007208}

\bibitem[{Konopliv {et~al.}(1999)Konopliv, Banerdt, \& Sjogren}]{Konopliv1999}
Konopliv, A., Banerdt, W., \& Sjogren, W. 1999, Icarus, 139, 3,
  \dodoi{10.1006/icar.1999.6086}

\bibitem[{Konopliv {et~al.}(1993)Konopliv, Borderies, Chodas, Christensen,
  Sjogren, Williams, Balmino, \& Barriot}]{Konopliv1993}
Konopliv, A.~S., Borderies, N.~J., Chodas, P.~W., {et~al.} 1993, Geophys. Res.
  Lett., 20, 2403, \dodoi{10.1029/93GL01890}

\bibitem[{Kopparapu {et~al.}(2014)Kopparapu, Ramirez, SchottelKotte, Kasting,
  Domagal-Goldman, \& Eymet}]{kop14}
Kopparapu, R., Ramirez, R., SchottelKotte, J., {et~al.} 2014, \apjl, 787, L29,
  \dodoi{10.1088/2041-8205/787/2/L29}

\bibitem[{Kopparapu {et~al.}(2013)Kopparapu, Ramirez, Kasting, Eymet, Robinson,
  Mahadevan, Terrien, Domagal-Goldman, Meadows, \& Deshpande}]{kop13}
Kopparapu, R., Ramirez, R., Kasting, J., {et~al.} 2013, \apj, 765, 131,
  \dodoi{10.1088/0004-637X/765/2/131}

\bibitem[{Korenaga(2013)}]{Korenaga2013}
Korenaga, J. 2013, Annu. Rev. Earth Planet. Sci., 41, 117,
  \dodoi{10.1146/annurev-earth-050212-124208}

\bibitem[{Lemoine {et~al.}(2001)Lemoine, Smith, Rowlands, Zuber, Neumann,
  Chinn, \& Pavlis}]{Lemoine2001}
Lemoine, F.~G., Smith, D.~E., Rowlands, D.~D., {et~al.} 2001, J. Geophys. Res.
  Planets, 106, 23359, \dodoi{10.1029/2000JE001426}

\bibitem[{{Magellan Team} {et~al.}(1997){Magellan Team}, Ford, Pettengill, Liu,
  \& Quigley}]{VenusDEM}
{Magellan Team}, Ford, P., Pettengill, G., Liu, F., \& Quigley, J. 1997,
  {Magellan Global Topography 4641m v02 (1997)}.
\newblock
  \url{https://astrogeology.usgs.gov/search/map/Venus/Magellan/RadarProperties/Venus_Magellan_Topography_Global_4641m_v02}

\bibitem[{McKenzie {et~al.}(1973)McKenzie, Roberts, \& Weiss}]{McKenzie1973}
McKenzie, D., Roberts, J., \& Weiss, N. 1973, Tectonophysics, 19, 89,
  \dodoi{10.1016/0040-1951(73)90034-6}

\bibitem[{McKenzie(1969)}]{McKenzie1969}
McKenzie, D.~P. 1969, Geophys. J. Int., 18, 1,
  \dodoi{10.1111/j.1365-246X.1969.tb00259.x}

\bibitem[{McKinney(2010)}]{mck10}
McKinney, W. 2010, in Proc. 9th Python Sci. Conf., ed. S.~van~der Walt \&
  J.~Millman, 56--61, \dodoi{10.25080/Majora-92bf1922-00a}

\bibitem[{McNamee {et~al.}(1993)McNamee, Borderies, \& Sjogren}]{McNamee1993}
McNamee, J.~B., Borderies, N.~J., \& Sjogren, W.~L. 1993, J. Geophys. Res.
  Planets, 98, 9113, \dodoi{10.1029/93JE00382}

\bibitem[{{Met Office}(2015)}]{Cartopy}
{Met Office}. 2015, {Cartopy: a cartographic python library with a Matplotlib
  interface}, Exeter, Devon.
\newblock \url{http://scitools.org.uk/cartopy}

\bibitem[{{MOLA Team}(2014)}]{MarsDEM}
{MOLA Team}. 2014, {Mars MGS MOLA Elevation Model 463m (MEGDR)}.
\newblock
  \url{https://astrogeology.usgs.gov/search/map/Mars/GlobalSurveyor/MOLA/Mars_MGS_MOLA_DEM_mosaic_global_463m}

\bibitem[{Moresi \& Solomatov(1998)}]{Moresi1998}
Moresi, L., \& Solomatov, V. 1998, Geophys. J. Int., 133, 669,
  \dodoi{10.1046/j.1365-246X.1998.00521.x}

\bibitem[{{National Imagery and Mapping Agency}(2000)}]{WGS84}
{National Imagery and Mapping Agency}. 2000, Tech. Rep., TR8350.2.
\newblock
  \url{http://earth-info.nga.mil/GandG/publications/tr8350.2/tr8350_2.html}

\bibitem[{Nearing {et~al.}(2004)Nearing, Pruski, \&
  O'Neal}]{nearing2004expected}
Nearing, M.~A., Pruski, F.~F., \& O'Neal, M.~R. 2004, J. soil water Conserv.,
  59, 43

\bibitem[{Neumann {et~al.}(2001)Neumann, Rowlands, Lemoine, Smith, \&
  Zuber}]{Neumann2001}
Neumann, G.~A., Rowlands, D.~D., Lemoine, F.~G., Smith, D.~E., \& Zuber, M.~T.
  2001, J. Geophys. Res. Planets, 106, 23753, \dodoi{10.1029/2000JE001381}

\bibitem[{Neumann {et~al.}(2003)Neumann, Smith, \& Zuber}]{Neumann2003}
Neumann, G.~A., Smith, D.~E., \& Zuber, M.~T. 2003, J. Geophys. Res., 108,
  5023, \dodoi{10.1029/2002JE001849}

\bibitem[{Nimmo \& McKenzie(1998)}]{Nimmo1998}
Nimmo, F., \& McKenzie, D. 1998, Annu. Rev. Earth Planet. Sci., 26, 23,
  \dodoi{10.1146/annurev.earth.26.1.23}

\bibitem[{Phillips(1986)}]{Phillips1986}
Phillips, R.~J. 1986, Geophys. Res. Lett., 13, 1141,
  \dodoi{10.1029/GL013i011p01141}

\bibitem[{Phillips {et~al.}(1992)Phillips, Raubertas, Arvidson, Sarkar,
  Herrick, Izenberg, \& Grimm}]{Phillips1992}
Phillips, R.~J., Raubertas, R.~F., Arvidson, R.~E., {et~al.} 1992, J. Geophys.
  Res. Planets, 97, 15923, \dodoi{10.1029/92JE01696}

\bibitem[{Ramirez(2018)}]{Ramirez2018}
Ramirez, R. 2018, Geosciences, 8, 280, \dodoi{10.3390/geosciences8080280}

\bibitem[{Ramirez \& Kaltenegger(2016)}]{Ramirez2016}
Ramirez, R., \& Kaltenegger, L. 2016, Astrophys. J., 823, 1,
  \dodoi{10.3847/0004-637X/823/1/6}

\bibitem[{Rapp(1989)}]{Rapp1989}
Rapp, R.~H. 1989, Geophys. J. Int., 99, 449,
  \dodoi{10.1111/j.1365-246X.1989.tb02031.x}

\bibitem[{Rappaport {et~al.}(1999)Rappaport, Konopliv, Kucinskas, \&
  Ford}]{Rappaport1999}
Rappaport, N.~J., Konopliv, A.~S., Kucinskas, A.~B., \& Ford, P.~G. 1999,
  Icarus, 139, 19, \dodoi{10.1006/icar.1999.6081}

\bibitem[{Reese {et~al.}(1999)Reese, Solomatov, \& Moresi}]{Reese1999}
Reese, C., Solomatov, V., \& Moresi, L.-N. 1999, Icarus, 139, 67,
  \dodoi{10.1006/icar.1999.6088}

\bibitem[{Ribas {et~al.}(2005)Ribas, Guinan, Gudel, \& Audard}]{Ribas2005}
Ribas, I., Guinan, E.~F., Gudel, M., \& Audard, M. 2005, Astrophys. J., 622,
  680, \dodoi{10.1086/427977}

\bibitem[{Rolf {et~al.}(2018)Rolf, Steinberger, Sruthi, \& Werner}]{Rolf2018}
Rolf, T., Steinberger, B., Sruthi, U., \& Werner, S. 2018, Icarus, 313, 107,
  \dodoi{10.1016/j.icarus.2018.05.014}

\bibitem[{Romeo(2013)}]{Romeo2013}
Romeo, I. 2013, Planet. Space Sci., 87, 157, \dodoi{10.1016/j.pss.2013.07.010}

\bibitem[{Romeo \& Turcotte(2010)}]{Romeo2010}
Romeo, I., \& Turcotte, D. 2010, Planet. Space Sci., 58, 1374,
  \dodoi{10.1016/j.pss.2010.05.022}

\bibitem[{Sagan(1976)}]{Sagan1976}
Sagan, C. 1976, Nature, 261, 31, \dodoi{10.1038/261031a0}

\bibitem[{Schaber {et~al.}(1992)Schaber, Strom, Moore, Soderblom, Kirk,
  Chadwick, Dawson, Gaddis, Boyce, \& Russell}]{Schaber1992}
Schaber, G.~G., Strom, R.~G., Moore, H.~J., {et~al.} 1992, J. Geophys. Res.,
  97, 13257, \dodoi{10.1029/92JE01246}

\bibitem[{Scotese \& Wright(2018)}]{Scotese2018}
Scotese, C.~R., \& Wright, N. 2018, {PALEOMAP Paleodigital Elevation Models
  (PaleoDEMS) for the Phanerozoic}, Tech. rep., Department of Earth \&
  Planetary Sciences, Northwestern University, Evanston, Illinois.
\newblock
  \url{http://www.earthbyte.org/paleodem-resource-scotese-and-wright-2018/}

\bibitem[{Shirey {et~al.}(2008)Shirey, Kamber, Whitehouse, Mueller, \&
  Basu}]{Shirey2008}
Shirey, S.~B., Kamber, B.~S., Whitehouse, M.~J., Mueller, P.~A., \& Basu, A.~R.
  2008, Spec. Pap. 440 When Did Plate Tectonics Begin Planet Earth?, 2440, 1,
  \dodoi{10.1130/2008.2440(01)}

\bibitem[{Shirey \& Richardson(2011)}]{Shirey2011}
Shirey, S.~B., \& Richardson, S.~H. 2011, Science (80-. )., 333, 434,
  \dodoi{10.1126/science.1206275}

\bibitem[{Sleep(1995)}]{Sleep1995}
Sleep, N.~H. 1995, Rev. Geophys., 33, 199, \dodoi{10.1029/95RG00126}

\bibitem[{Smith {et~al.}(2001)Smith, Zuber, Frey, Garvin, Head, Muhleman,
  Pettengill, Phillips, Solomon, Zwally, Banerdt, Duxbury, Golombek, Lemoine,
  Neumann, Rowlands, Aharonson, Ford, Ivanov, Johnson, McGovern, Abshire,
  Afzal, \& Sun}]{Smith2001}
Smith, D.~E., Zuber, M.~T., Frey, H.~V., {et~al.} 2001, J. Geophys. Res.
  Planets, 106, 23689, \dodoi{10.1029/2000JE001364}

\bibitem[{Smrekar {et~al.}(2020)Smrekar, Dyar, Helbert, Hensley, Nunes, \&
  Whitten}]{Smrekar2020}
Smrekar, S., Dyar, D., Helbert, J., {et~al.} 2020, in Eur. Planet. Sci. Congr.,
  EPSC2020--447, \dodoi{10.5194/epsc2020-447}

\bibitem[{Smrekar {et~al.}(2018)Smrekar, Davaille, \& Sotin}]{Smrekar2018}
Smrekar, S.~E., Davaille, A., \& Sotin, C. 2018, Space Sci. Rev., 214, 88,
  \dodoi{10.1007/s11214-018-0518-1}

\bibitem[{Solomatov \& Moresi(1996)}]{Solomatov1996}
Solomatov, V.~S., \& Moresi, L.-N. 1996, J. Geophys. Res. Planets, 101, 4737,
  \dodoi{10.1029/95JE03361}

\bibitem[{Speagle(2019)}]{Speagle2020}
Speagle, J.~S. 2019, Mon. Not. R. Astron. Soc., 493, 3132,
  \dodoi{10.1093/mnras/staa278}

\bibitem[{Stein {et~al.}(2010)Stein, Fahl, \& Hansen}]{Stein2010}
Stein, C., Fahl, A., \& Hansen, U. 2010, Geophys. Res. Lett., 37, n/a,
  \dodoi{10.1029/2009GL041073}

\bibitem[{Stern(2016)}]{Stern2016}
Stern, R.~J. 2016, Geosci. Front., 7, 573, \dodoi{10.1016/j.gsf.2015.12.002}

\bibitem[{Stern {et~al.}(2018)Stern, Gerya, \& Tackley}]{Stern2018a}
Stern, R.~J., Gerya, T., \& Tackley, P.~J. 2018, Geosci. Front., 9, 103,
  \dodoi{10.1016/j.gsf.2017.06.004}

\bibitem[{Strom(1993)}]{Strom1993}
Strom, R. 1993, in Lunar and Planetary Science Conference, Vol.~24, Lunar
  Planet. Sci. Conf.

\bibitem[{Strom {et~al.}(1994)Strom, Schaber, \& Dawson}]{Strom1994}
Strom, R.~G., Schaber, G.~G., \& Dawson, D.~D. 1994, J. Geophys. Res., 99,
  10899, \dodoi{10.1029/94JE00388}

\bibitem[{Turcotte(1987)}]{Turcotte1987}
Turcotte, D.~L. 1987, J. Geophys. Res. Solid Earth, 92, E597,
  \dodoi{10.1029/JB092iB04p0E597}

\bibitem[{Turcotte(1993)}]{Turcotte1993}
---. 1993, J. Geophys. Res., 98, 17061, \dodoi{10.1029/93JE01775}

\bibitem[{Turcotte(1995)}]{Turcotte1995}
---. 1995, J. Geophys. Res., 100, 16931, \dodoi{10.1029/95JE01621}

\bibitem[{van~der Walt {et~al.}(2011)van~der Walt, Colbert, \&
  Varoquaux}]{van11}
van~der Walt, S., Colbert, S.~C., \& Varoquaux, G. 2011, Comput. Sci. Eng., 13,
  22, \dodoi{10.1109/MCSE.2011.37}

\bibitem[{van Hunen \& Moyen(2012)}]{vanHunen2012}
van Hunen, J., \& Moyen, J.-F. 2012, Annu. Rev. Earth Planet. Sci., 40, 195,
  \dodoi{10.1146/annurev-earth-042711-105255}

\bibitem[{Vening-Meinesz(1951)}]{Vening-Meinesz1951}
Vening-Meinesz, F.~A. 1951, Proc. K. Ned. Acad. van Wet. Ser. B Phys. Sci., 1

\bibitem[{Way \& {Del Genio}(2020)}]{Way2020}
Way, M.~J., \& {Del Genio}, A.~D. 2020, J. Geophys. Res. Planets, 125,
  \dodoi{10.1029/2019JE006276}

\bibitem[{Way {et~al.}(2022)Way, Ernst, \& Scargle}]{Way2022}
Way, M.~J., Ernst, R.~E., \& Scargle, J.~D. 2022, Planet. Sci. J., 3, 92,
  \dodoi{10.3847/PSJ/ac6033}

\bibitem[{Weller {et~al.}(2015)Weller, Lenardic, \& O'Neill}]{Weller2015}
Weller, M., Lenardic, A., \& O'Neill, C. 2015, Earth Planet. Sci. Lett., 420,
  85, \dodoi{10.1016/j.epsl.2015.03.021}

\bibitem[{Weller {et~al.}(2023)Weller, Evans, Ibarra, \& Johnson}]{Weller2023}
Weller, M.~B., Evans, A.~J., Ibarra, D.~E., \& Johnson, A.~V. 2023, Nat.
  Astron., \dodoi{10.1038/s41550-023-02102-w}

\bibitem[{Weller \& Kiefer(2020)}]{Weller2020}
Weller, M.~B., \& Kiefer, W.~S. 2020, J. Geophys. Res. Planets, 125, 1,
  \dodoi{10.1029/2019JE005960}

\bibitem[{Wessel \& M{\"{u}}ller(2007)}]{Wessel2007}
Wessel, P., \& M{\"{u}}ller, R. 2007, Treatise Geophys., Part 2, 49,
  \dodoi{10.1016/B978-044452748-6.00101-2}

\bibitem[{Whitehead(2017)}]{Whitehead2017}
Whitehead, J. 2017, Earth Planet. Sci. Lett., 467, 18,
  \dodoi{10.1016/j.epsl.2017.03.017}

\bibitem[{Wieczorek \& Meschede(2018)}]{Wieczorek2018}
Wieczorek, M.~A., \& Meschede, M. 2018, Geochemistry, Geophys. Geosystems, 19,
  2574, \dodoi{10.1029/2018GC007529}

\bibitem[{Wilson(1965)}]{Wilson1965}
Wilson, J.~T. 1965, Nature, 207, 343, \dodoi{10.1038/207343a0}

\end{thebibliography}
\bibliographystyle{aasjournal}

\end{document}